\documentclass[12pt]{article}
\usepackage{a4wide}
\usepackage{amssymb}
\usepackage{graphicx}
\begin{document}
{\renewcommand{\thefootnote}{\fnsymbol{footnote}}
\begin{center}
{\LARGE  Non-covariance of ``covariant polymerization''\\ in models of
  loop quantum gravity}\\
\vspace{1.5em}
Martin Bojowald\footnote{e-mail address: {\tt bojowald@gravity.psu.edu}}
\\
\vspace{0.5em}
Institute for Gravitation and the Cosmos,\\
The Pennsylvania State
University,\\
104 Davey Lab, University Park, PA 16802, USA\\
\vspace{1.5em}
\end{center}
}

\setcounter{footnote}{0}

\begin{abstract}
  It is possible to implement a certain form of modified gravity inspired by
  loop quantization through non-bijective canonical transformations. The
  canonical nature might suggest that such modifications are guaranteed to
  preserve general covariance. Here, however, we show that a dedicated
  space-time analysis is still required, even in the case of a bijective
  canonical transformation. In addition, a complete global analysis is
  presented for a recent proposal of a non-bijective transformation, showing
  that it does not preserve general covariance and that the only novel
  physical effect introduced by the modification is the presence of certain
  time-reversal hypersurfaces between classical space-time regions. These
  results provide further insights into the physical interpretation of
  modified dynamics in models of loop quantum gravity.
\end{abstract}

\section{Introduction}

Models of loop quantum gravity attempt to implement quantum-geometry effects
by using certain modifications of the classical equations of canonical
gravity. The canonical nature, as usual, implies that general covariance is
not manifest and must be tested by dedicated means. Several no-go results for
general covariance and slicing independence in such models have recently been
derived, using setups relevant for cosmology \cite{NonCovDressed} and
black-holes \cite{Disfig,BlackHoleModels}. The only known way to realize
covariance in models of loop quantum gravity is through a deformed version
\cite{Action,DeformedRel} that implies signature change at high density or
curvature when applied to modifications commonly used in loop quantum
cosmology or loop quantum black holes
\cite{Action,SigChange,SigImpl,SigChangeEmerge,BHSigChange,DefSchwarzschild,DefSchwarzschild2,DefGenBH}. (Signature
change may be avoided in some cases, but it would require non-standard
modifications such as complex connections
\cite{CosmoComplex,SphSymmComplex,GowdyComplex,DeformSpinComplex},
Euclidean-type gravity \cite{LQCScalarEucl,EuclConn} or non-bouncing
background solutions \cite{NonBouncing}.)

It is therefore important to explore possible alternative modifications. In
this context, \cite{CovPol} suggest to apply a non-bijective canonical
transformation to the classical theory, hoping that the modified model will be
close enough to the classical system to preserve covariance, yet different
enough to be considered a modification because the transformation is not
bijective. As we will show in this paper, covariance is a subtle issue even in
this case and must be derived. Once this task has been completed, it can be
seen that the modifications are not compatible with general covariance or
slicing independence in a global space-time structure. The equations suggested
in \cite{CovPol} therefore do not show how models of loop quantum gravity
could be made consistent with general covariance, and they do not provide
counter-examples to the no-go results of \cite{NonCovDressed,Disfig}.

Our analysis of general covariance makes use of effective line elements, as
defined in \cite{EffLine}.  A proper effective line element provides a
geometrical interpretation of solutions of a modified theory of gravity. For
the line element to have a proper geometrical meaning, it must be invariant
under coordinate changes. But modified equations of a model may well change
the gauge transformations imposed on basic fields, in particular if the model
is formulated canonically and does not make use of space-time
tensors. Therefore, the existence of suitable metric components constructed
from the basic fields of the modified theory such that they form an invariant
line element is, in general, not guaranteed. Even if metric components exist,
their relationship with the basic fields is usually modified, compared with
the classical relationship, in order to account for modified gauge
transformations.

In \cite{CovPol} and elsewhere in the literature, however, the simple
classical relationship between metric components and basic fields is
mistakenly assumed to hold also in the presence of modifications. A derivation
of proper effective line elements then corrects the resulting understanding of
space-time structure, and it reveals the global geometry implied by solutions
of the modified theory. As a result, solutions of \cite{CovPol} are simply
concatenations of classical space-time regions, separated by time-reversal
hypersurfaces. These hypersurfaces, derived in more detail in
Section~\ref{s:Global} below, are implicitly defined by time derivatives of
canonical fields changing sign in a discontinuous manner.  Their presence
makes it possible for extrinsic curvature to remain bounded. However, they can
be defined only using non-invariant quantities, thus violating covariance on a
global level.

In addition to the suggestion made in \cite{CovPol}, we will also consider the
case of a bijective canonical transformation. Such a transformation should, of
course, exactly preserve physical properties of the classical theory,
including general covariance. Nevertheless, we will see that space-time
structure in such a ``modified'' canonical theory is non-trivial and requires
a dedicated analysis before physical conclusions can be drawn. The model
therefore provides an instructive example: Even though it is unable to imply
new physics, a careless analysis might wrongly suggest new effects such as
singularity resolution. These lessons will then be applied to the model
proposed in \cite{CovPol}. They are also relevant more broadly in a large
number of models of loop quantum gravity in which line elements have been used
for modified theories without confirming their geometrical validity
\cite{BlackHoleModels}.

The results of the present paper demonstrate the importance of considering
properly defined effective line elements to express solutions of equations of
motion in modified canonical theories of gravity. They also underline the
highly non-trivial nature of covariance in models of loop quantum gravity,
which turns out to be violated even by the minimal modifications suggested in
\cite{CovPol}, based on a canonical transformation from the classical theory.

\section{Space-time analysis}

We present a detailed space-time analysis of the model introduced in
\cite{CovPol}. Since the model is canonical, we use methods of canonical
gravity; see \cite{CUP,Foundations} for details. 

\subsection{Variables and transformations}

Canonical gravity of spherically symmetric models is described by line
elements of the form \cite{ADM}
\begin{equation} \label{ds}
 {\rm d}s^2= -N^2{\rm d}t^2+ q_{xx} ({\rm d}x+M{\rm d}t)^2+ q_{\varphi\varphi}
 ({\rm d}\vartheta^2+\sin^2\vartheta{\rm d}\varphi)\,.
\end{equation}
The spatial part is determined by two functions, $q_{xx}$ and
$q_{\varphi\varphi}$, depending on the radial position $x$ as well as time
$t$, while the lapse function $N$ and shift vector $M$, also depending on $x$
and $t$, describe its extension to space-time. In spherically symmetric models
of loop quantum gravity \cite{SymmRed,SphSymmHam}, one usually replaces metric
components with components $E^x$ and $E^{\varphi}$ of a densitized triad, such
that
\begin{equation} \label{qE}
 q_{xx}= \frac{(E^{\varphi})^2}{|E^x|} \quad,\quad q_{\varphi\varphi}=|E^x|\,.
\end{equation}
In what follows it will be sufficient to assume $E^x>0$, fixing the
orientation of space. 

The triad components are, up to constant factors, canonically conjugate to
components of extrinsic curvature, $K_x$ and $K_{\varphi}$, such that
\begin{equation}
 \{K_x(x_1),E^x(x_2)\}=2G\delta(x_1,x_2) \quad,\quad
 \{K_{\varphi}(x_1),E^{\varphi}(x_2)\}=G\delta(x_1,x_2) 
\end{equation}
with Newton's constant $G$. Extrinsic curvature depends on time and space
derivatives of the densitized triad (as well as lapse and shift) in a way that
may be modified in models of loop quantum gravity. We will not need the
precise relationships but only use the canonical structure.

Depending on the time gauge, equations of motion for the basic phase-space
variables are generated by combinations of the Hamiltonian constraint, $H[N]$,
and the diffeomorphism constraint, $D[M]$. We will not need the precise form
of these expressions either but only refer to their nature as gauge generators
of deformations of spatial hypersurfaces in space-time. These transformations
correspond to classical space-time \cite{Regained} provided the constraints
obey Dirac's hypersurface-deformation brackets \cite{DiracHamGR}, in
particular
\begin{equation} \label{HH}
 \{H[N_1],H[N_2]\}= -D[E^x(E^{\varphi})^{-2} (N_1N_2'-N_1'N_2)]\,.
\end{equation}
The presence of a phase-space dependent structure function implies that the
structure of space-time is sensitive to modifications of the constraints. 

As shown in \cite{LoopSchwarz}, the structure function can be eliminated in an
equivalent constrained system obtained by suitable combinations of $H$ and
$D$. This construction has also been used in the recent analysis of
\cite{CovPol}. However, based on \cite{Regained}, the behavior of hypersurface
deformations and therefore of general covariance and slicing independence
requires a bracket of the form (\ref{HH}) for the generators of normal
deformations of spatial hypersurfaces. Discussions of covariance therefore
cannot avoid referring to this relationship, especially in attempted
modifications.

The main ingredient in models of loop quantum gravity is a substitution of
(almost) periodic functions of connection or extrinsic-curvature components
for the classical quadratic dependence in the Hamiltonian constraint. If this
substitution is done only in these places, and in a careful way relating
different substitution functions to one another, the bracket (\ref{HH}) in
vacuum is modified by a new factor of the structure function such that the
structure of space-time is non-classical
\cite{JR,HigherSpatial,SphSymmOp}. (See \cite{ScalarHolInv,DeformedCosmo} for
an analogous result in the cosmological context.) In the presence of a scalar
field, no such substitution is known that preserves the form of (\ref{HH})
even if one accepts modifications of the structure function \cite{SphSymmCov}.

The authors of \cite{CovPol} suggest that this difficulty may be overcome if
one uses a canonical transformation instead of substitution. For the
gravitational variables, they propose to transform from the pair
$(K_{\varphi},E^{\varphi})$ to a new pair
$(\tilde{K}_{\varphi},\tilde{E}^{\varphi})$ such that
\begin{equation} \label{Can}
 K_{\varphi}= \frac{\sin(\delta \tilde{K}_{\varphi})}{\delta} \quad,\quad
 E^{\varphi}= \frac{\tilde{E}^{\varphi}}{\cos(\delta \tilde{K}_{\varphi})}\,.
\end{equation}
The pair $(K_x,E^x)$ remains unchanged. There is a similar transformation for
a scalar matter field, which we do not use explicitly here because (\ref{Can})
is sufficient for a discussion of space-time structure: The scalar field does
not appear in the structure function of (\ref{HH}).

Expressed in terms of the new variables, the Hamiltonian constraint depends on
$\tilde{K}_{\varphi}$ through a periodic function, as in standard
modifications, while the dependence of $E^{\varphi}$ on $\tilde{K}_{\varphi}$
leads to new modifications in metric functions not considered before. The hope
is that these new modifications may preserve general covariance because the
model is obtained by a canonical transformation from a covariant theory. At
the same time, only a bounded range of $K_{\varphi}$ is realized for an
infinite range of $\tilde{K}_{\varphi}$, which could introduce new physical
effects and help with the resolution of singularities.

\subsection{Bijective canonical transformation}

The model of \cite{CovPol} is based on a canonical transformation of the
classical theory which is not bijective, and therefore need not be completely
equivalent to classical gravity. It may therefore be considered a modified
version of spherically symmetric general relativity. The case of a bijective
canonical transformation, by contrast, could be deemed too trivial to be
worthy of attention because it cannot lead to new physics. It is nevertheless
instructive to see how a dedicated space-time analysis would proceed if we
were faced with a proposed modified theory without knowing that it is simply
obtained by a bijective canonical transformation from classical general
relativity.

The setup is therefore as follows: We are given a canonical theory with
canonical pairs $(\tilde{K}_{\varphi},\tilde{E}^{\varphi})$ and $(K_x,E^x)$
and perhaps some matter fields, as well as a consistent set of diffeomorphism
and Hamiltonian constraints in these variables.  The consistent constraints
have been derived by applying a bijective canonical transformation
\begin{equation} \label{Canf}
 K_{\varphi} = f(\tilde{K}_{\varphi}) \quad,\quad
 E^{\varphi}=\frac{\tilde{E}^{\varphi}}{{\rm d}f/{\rm d}\tilde{K}_{\varphi}}
\end{equation}
to the constraints of classical spherically symmetric gravity in canonical
form, where $f$ is a monotonic function such that
$f(\tilde{K}_{\varphi})\approx \tilde{K}_{\varphi}$ for $\tilde{K}_{\varphi}$
sufficiently small compared with some reference scale. Given these conditions,
$f$ may well be such that the full range of $K_{\varphi}$ is mapped to a
finite range of $\tilde{K}_{\varphi}$ in which case the transformation would
be bijective provided the new variable $\tilde{K}_{\varphi}$ is always
restricted to this finite range.  In spite of the underlying equivalence with
classical gravity, one could therefore claim that consistent constraints imply
new physics and that singularities are resolved because curvature
($\tilde{K}_{\varphi}$) remains bounded, all while preserving general
covariance.

More generally, we could assume a bijective 2-variable transformation
\begin{equation} \label{Canf2}
 K_{\varphi} = f_1(\tilde{K}_{\varphi},\tilde{E}^{\varphi}) \quad,\quad E^{\varphi} =
 f_2(\tilde{K}_{\varphi},\tilde{E}^{\varphi})
\end{equation}
such that $\{f_1,f_2\}=G$. It would not be straightforward to reconstruct this
transformation if we were just given the resulting constraints. How would we
then spot possible erroneous claims of new physics and show that the theory
is, in fact, completely equivalent to spherically symmetric general
relativity?

To some extent, the situation is comparable to the task of telling that a
``new'' solution of general relativity has just been obtained from a
well-known one by a coordinate transformation. Like a canonical
transformation, a coordinate transformation, if incompletely analyzed, could
also suggest bounded curvature if it maps a finite space-time region that does
not include singularities into a full infinite range of a new coordinate. In
this case, there are standard methods to analyze the global meaning of
solutions, for instance by checking geodesic completeness to determine whether
an infinite range of some coordinate amounts to an infinite geometric
distance, or just to some finite interval.

At this point, however, the two examples of a canonical transformation and a
coordinate transformation start to differ conceptually. While any coordinate
transformation preserves space-time structure and covariance, a canonical
transformation need not do so. In particular, a coordinate transformation
gives us an unambiguous new metric to be used for a geometrical
derivation. But a canonical transformation, without further analysis, does not
tell us whether some new field $\tilde{E}^{\varphi}$ can indeed be used in a
metric component just like the original $E^{\varphi}$, or whether the new
$\tilde{K}_{\varphi}$ is indeed a curvature component with the same
geometrical meaning as $K_{\varphi}$. At this point, at the latest, we should
become suspicious of claims about eliminated singularities in a bijectively
transformed theory because a bounded $\tilde{K}_{\varphi}$ does not
necessarily imply bounded curvature. How do we turn our suspicion into a proof
that the singularity claims are incorrect?

\subsection{Effective line elements}
\label{s:EffLine}

A canonical space-time analysis gives us a clear answer to the questions posed
in the preceding subsection.  Solutions of a modified canonical theory of
gravity are not necessarily geometrical, that is, one cannot simply assume
that inserting some $\tilde{E}^{\varphi}$ instead of $E^{\varphi}$ in
(\ref{qE}) results in a well-defined space-time line element of the form
(\ref{ds}) with the same lapse $N$ and shift $M$ as used in the relevant
equations of motion. Any line element ${\rm d}s^2=g_{\alpha\beta}{\rm
  d}x^{\alpha}{\rm d}x^{\beta}$, by definition, has to be invariant with
respect to a combination of coordinate transformations of ${\rm d}x^{\alpha}$
and gauge transformations of the canonical metric components.

While ${\rm d}x$ and ${\rm d}t$ in (\ref{ds}) still transform like standard
coordinate differentials after applying a canonical transformation such as
(\ref{Can}), (\ref{Canf}) or (\ref{Canf2}), the new field
$\tilde{E}^{\varphi}$ does not have the same (gauge) transformation behavior
as the classical $E^{\varphi}$ because the transformation depends on
$\tilde{K}_{\varphi}$ which, like $K_{\varphi}$, is not a space-time
scalar. Therefore, using a modified $\tilde{E}^{\varphi}$ in $q_{xx}$ for
(\ref{ds}) implies that modified metric components no longer transform in a
way dual to coordinate differentials, and the line element is not
invariant. Geometrical derivations from such an expression are meaningless
because they depend on coordinate choices. (One could try to modify the
transformations of ${\rm d}x$ and ${\rm d}t$ to compensate for the modified
gauge transformations of $\tilde{E}^{\varphi}$, for instance by using
non-classical manifolds. However, no such manifold structure is known for the
specific modifications discussed here. For the example of non-commutative
manifolds from the perspective of hypersurface deformations, see
\cite{NCHDA}.)

As shown in \cite{EffLine}, it is sometimes possible to apply a field
redefinition to canonical fields in a modified theory so as to bring their
gauge transformations to a form required for an invariant effective line
element. In the present case, one can use methods introduced in \cite{Absorb}
to find a suitable field redefinition of $\tilde{E}^{\varphi}$, which can be
summarized as follows: A field $E^{\varphi}$ that, together with its conjugate
$K_{\varphi}$, appears in the Hamiltonian and diffeomorphism constraints of a
canonical theory plays the role of a metric component as in (\ref{qE}) if and
only if the Poisson bracket of two Hamiltonian constraints equals
(\ref{HH}). In the bijectively transformed theory, however, this bracket is
replaced by
\begin{equation} \label{HHmodf2}
\{H[N_1],H[N_2]\}=
 -D[E^xf_2(\tilde{K}_{\varphi},\tilde{E}^{\varphi})^{-2}
 (N_1N_2'-N_1'N_2)]\,,
\end{equation}
or
\begin{equation} \label{HHmodf}
\{H[N_1],H[N_2]\}=
 -D[({\rm d}f/{\rm d}\tilde{K}_{\varphi})^{-2}
 E^x(\tilde{E}^{\varphi})^{-2} 
 (N_1N_2'-N_1'N_2)]
\end{equation}
in the simpler 1-variable transformation. Therefore, using
$\tilde{E}^{\varphi}$ in (\ref{qE}) does not yield a legitimate metric
component, and $\tilde{K}_{\varphi}$ is not a component of extrinsic
curvature.

In order to derive the correct space-time structure and a meaningful metric,
we should find a suitable function $\tilde{\tilde{E}}^{\varphi}$ of
$\tilde{E}^{\varphi}$ and $\tilde{K}_{\varphi}$ in terms of which the Poisson
bracket of two Hamiltonian constraints takes on the classical form
(\ref{HH}). It is easy to see that $\tilde{\tilde{E}}^{\varphi}=E^{\varphi}$
is just the classical field in (\ref{HHmodf2}) or (\ref{HHmodf}). Completing
this substitution to a canonical transformation then leads us back to the
classical $K_{\varphi}$ from $\tilde{K}_{\varphi}$, and inserting this
transformation in the constraints tells us that the theory is nothing but
classical. (The canonical conjugate $K_{\varphi}$ of some function
$E^{\varphi}$ on the phase space $(\tilde{K}_{\varphi},\tilde{E}^{\varphi})$
is not uniquely determined because any function of $E^{\varphi}$ could be
added to $K_{\varphi}$ while maintaining the nature of a canonical
conjugate. However, this freedom is eliminated by the boundary condition that
$K_{\varphi}\approx \tilde{K}_{\varphi}$ for $\tilde{K}_{\varphi}$ small with
respect to some scale used in the model.)  At this point, we would have
debunked any potential claims of new physics and singularity resolution.

Our example is artificial and deals with a trivial modification of classical
general relativity. It is nevertheless instructive because it shows the
importance of a dedicated analysis of space-time structure in canonical
terms. It is also relevant because arguments comparable to some ingredients of
our example have often been made in models of loop quantum gravity. These
models deal with actual modifications of classical gravity and there is a
possibility for new physics to emerge. But also in this case, it is often, and
incorrectly, assumed that some field $\tilde{E}^{\varphi}$ that shows some
semblence to the classical $E^{\varphi}$ can be used to define a meaningful
metric component using (\ref{qE}). This geometrical interpretation is possible
only if $\tilde{E}^{\varphi}$ is such that the Poisson bracket of two
Hamiltonian constraints equals (\ref{HH}) where $E^{\varphi}$ is simply
replaced by $\tilde{E}^{\varphi}$, without introducing any multiplicative
factor or other modifications of the structure function. Unfortunately, this
condition is rarely realized in models of loop quantum gravity, which often do
not even check that the Poisson bracket of two Hamiltonian constraints remains
closed after modifications.

\section{Polymerized models}

In the case of \cite{CovPol}, it is clear that the bracket of two Hamiltonian
constraints remains closed after applying a non-bijective canonical
transformation. Moreover, the modification is non-trivial because the
canonical transformation used in this case, given by (\ref{Can}), is not
bijective. As seen in the preceding subsection, however, a dedicated
space-time analysis is necessary to interpret the theory even in the case of a
bijective transformation. It should then certainly be performed also in the
non-bijective case, but this has not been attempted in \cite{CovPol}. It is
therefore unclear whether physical statements suggested there are correct.

The modified theory has Hamiltonian constraints such that
\begin{equation} \label{HHmod}
 \{H[N_1],H[N_2]\}=
 -D[\cos^2(\delta\tilde{K}_{\varphi})E^x(\tilde{E}^{\varphi})^{-2}
 (N_1N_2'-N_1'N_2)]\,. 
\end{equation}
with a modified structure function, obtained by simply applying the canonical
transformation to (\ref{HH}). Since the modification introduces new zeros of
the structure function at $\delta\tilde{K}_{\varphi}=\frac{1}{2}(2n+1)\pi$
with integer $n$, it eliminates some contributions of the diffeomorphism
constraint from the right-hand side. The presence of structure functions
implies that generators of hypersurface deformations form a Lie algebroid
\cite{Pradines,Weinstein,ConsAlgebroid} over phase space, labeling
independent contributions from the constraints. New zeros in the structure
function introduced by the transformation mean that the algebroid gains new
Abelian subalgebroids by restriction to the zero-level sets of the structure
function. The algebraic structure is therefore inequivalent to its classical
form. (The authors of \cite{CovPol} claim that the modification ``preserves
the constraint algebra,'' which presumably refers to a partial Abelianization
of the generators as in \cite{LoopSchwarz}. However, as shown in
\cite{SphSymmCov}, such a reformulation of the constraints is not sufficient
for a discussion of general covariance.)

An inequivalent algebraic structure of hypersurface deformations implies that
covariance is non-trivial in the modified system. As the authors of
\cite{CovPol} point out, the canonical transformation employed to obtain the
modification is not bijective. This property is the reason why there are
additional zeros in the modified structure function of
hypersurface-deformation brackets. According to \cite{CovPol}, the
non-bijective nature of the transformation might provide a chance for the
modified theory to describe new physical effects, but it is also the reason
why covariance is no longer obvious even though the modification has been
obtained by canonically transforming a covariant theory. The claim ``It has
the advantage that it is a canonical transformation from the original
variables. That means that it preserves the constraint algebra and the
covariance of the theory, which previous choices did not.'' of \cite{CovPol}
is therefore incorrect. In the presence of modified hypersurface deformations
with an inequivalent algebraic structure, covariance has to be derived by a
careful analysis of generic solutions and their geometrical meaning.

\subsection{Local solutions}

Local solutions for $\tilde{E}^{\varphi}$ and $\tilde{K}_{\varphi}$ can be
derived without explicitly solving modified equations of motion because they
can simply be obtained by applying a local (in phase space) inverse of the
canonical transformation (\ref{Can}) to a classical solution in canonical
form. Starting at small $\delta K_{\varphi}$ for the classical solution, any
modified local solution $\tilde{K}_{\varphi}$ remains valid until $\delta
K_{\varphi}$ reaches the values $\pm 1$, the local maxima of $\sin(\delta
\tilde{K}_{\varphi})$ where the canonical transformation is no longer
invertible.

If one were to solve modified equations directly for
$(\tilde{K}_{\varphi},\tilde{E}^{\varphi})$, starting with some initial
values, it would be possible to cross regions where $\delta
\tilde{K}_{\varphi}=\pm \frac{1}{2}\pi$, again corresponding to the first
local maxima of $\sin(\delta \tilde{K}_{\varphi})$ close to small $\delta
K_{\varphi}$. Such an extension of the local solution is no longer a simple
local inverse of the canonical transformation, and presumably gives rise to
``novel phenomena'' that are, according to \cite{CovPol}, introduced by the
modification.

However, a solution in the range where $\delta \tilde{K}_{\varphi}>
\frac{1}{2}\pi$ (the case of $\delta \tilde{K}_{\varphi}<- \frac{1}{2}\pi$
being analogous) and $\delta \tilde{K}_{\varphi}<\frac{3}{2}\pi$, can again be
interpreted as a local inverse of (\ref{Can}), but one that makes use of a
different branch of the arcsine compared with the initial region at $|\delta
\tilde{K}_{\varphi}|< \frac{1}{2}\pi$. The canonical transformation therefore
provides a classical analog in any range of $\delta \tilde{K}_{\varphi}$ that
excludes the values $\frac{1}{2}(2n+1)\pi$ with integer $n$. While the
analogous $K_{\varphi}$ is always bounded thanks to (\ref{Can}), there is no
upper limit on $\delta \tilde{K}_{\varphi}$ beyond which classical analogs
would no longer exist.

We have obtained a direct correspondence between local solutions in the
classical and modified theories. The next question we have to address is
whether physics or geometry in the modified theory should be based on the
field $\tilde{K}_{\varphi}$ and its conjugate $\tilde{E}^{\varphi}$, or on
their local classical analogs $K_{\varphi}$ and $E^{\varphi}$. This question
is relevant for the application presented in \cite{CovPol}, in which critical
collapse is studied numerically by evaluating a ``black hole mass.''
Unfortunately, \cite{CovPol} does not specify how this mass is obtained, but
presumably it refers to a mass parameter extracted in the usual way from a
line element, constructed from $\tilde{E}^{\varphi}$ rather than $E^{\varphi}$
in the modified theory. We therefore have to analyze how a meaningful line
element can be constructed in the modified theory. A meaningful line element
requires specific transformation properties to hold for its coefficients.

\subsection{Effective line elements}

Using local inverses of the canonical transformation, we have obtained local
solutions in canonical form, resulting in evolutions of $\tilde{K}_{\varphi}$
and $\tilde{E}^{\varphi}$ depending on some time coordinate implicitly
determined by lapse and shift. Such a solution of equations of motion in a
modified theory is not necessarily geometrical, that is, one cannot simply
assume that inserting $\tilde{E}^{\varphi}$ instead of $E^{\varphi}$ in
(\ref{qE}) results in a well-defined space-time line element of the form
(\ref{ds}) with the same lapse $N$ and shift $M$ as used in the relevant
equations of motion. Any line element, by definition, has to be invariant with
respect to a combination of coordinate transformations of ${\rm d}x^{\alpha}$
and gauge transformations of the canonical metric components.

While ${\rm d}x$ and ${\rm d}t$ still transform like standard coordinate
differentials after applying a canonical transformation such as (\ref{Can}),
the new field $\tilde{E}^{\varphi}$ does not have the same (gauge)
transformation behavior as the classical $E^{\varphi}$ because $K_{\varphi}$
in (\ref{Can}) is not a space-time scalar. Therefore, using a modified
$\tilde{E}^{\varphi}$ in $q_{xx}$ for (\ref{ds}) implies that modified metric
components no longer transform in a way dual to coordinate differentials, and
the line element is not invariant. Geometrical derivations from such an
expression are meaningless because they depend on coordinate choices. (One
could try to modify the transformations of ${\rm d}x$ and ${\rm d}t$ to
compensate for the modified gauge transformations of $\tilde{E}^{\varphi}$,
for instance by using non-classical manifolds. However, no such manifold
structure is known for the specific modifications discussed here. For the
example of non-commutative manifolds from the perspective of hypersurface
deformations, see \cite{NCHDA}.)

As shown in \cite{EffLine}, it is sometimes possible to apply a field
redefinition to canonical fields in a modified theory so as to bring their
gauge transformations to a form required for an invariant effective line
element. In the present case, one can use methods introduced in \cite{Absorb}
to find a suitable field redefinition of $\tilde{E}^{\varphi}$. Not
surprisingly, this field redefinition is simply an application of the
canonical transformation (\ref{Can}), mapping $\tilde{E}^{\varphi}$ back to
$E^{\varphi}$ which clearly has the correct transformation behavior for a
well-defined line element.

Methods of effective line elements therefore show that physics and geometry in
the modified theory should be based on the classical analogs found in the
previous subsection, and not on the modified solutions $\tilde{K}_{\varphi}$
and $\tilde{E}^{\varphi}$. In any region in which (\ref{Can}) is locally
invertible, the modified theory simply describes a transformed version of
classical gravity. Any potential for new physical effects is restricted to
subsets of measure zero in phase space and (generically) space-time. In order
to understand their meaning, we have to determine how different regions of
classical analogs may be connected in an effective space-time picture of
global form.

\subsection{Global structure}
\label{s:Global}

So far, we have obtained formal piecewise solutions for the canonical fields
$\tilde{K}_{\varphi}$ and $\tilde{E}^{\varphi}$ as well as effective line
elements that faithfully describe their geometrical meaning, based on field
redefinitions. The final question is how these piecewise solutions can be
glued back together to obtain a global space-time picture. Such a gluing
cannot be based on classical matching conditions because they would simply
lead to a global classical solution that does not respect the boundedness of
$K_{\varphi}$ implied by (\ref{Can}).

Given a solution for $\tilde{K}_{\varphi}$ and $\tilde{E}^{\varphi}$, a
classical analog and an effective line element is obtained by applying the
canonical transformation (\ref{Can}). Since the transformation is not
bijective, different ranges of $\tilde{K}_{\varphi}$ may correspond to the
same classical geometry. If we first restrict ourselves to ranges of
$\tilde{K}_{\varphi}$ in which the transformation is invertible, the
corresponding phase-space region corresponds, via the effective line element,
to a region in space-time which generically is incomplete because it is cut
off at fixed values of $K_{\varphi}$. A global solution therefore requires an
extension through the hypersurfaces on which
$\delta\tilde{K}_{\varphi}=\frac{1}{2}(2n+1)\pi$ with integer $n$.

It is easy to see how different regions are connected if we first focus on two
neighbors, such as the low-curvature region, called region I where
$|\delta\tilde{K}_{\varphi}|<\frac{1}{2}\pi$, and a region II where
$\frac{1}{2}\pi<\delta\tilde{K}_{\varphi}<\frac{3}{2}\pi$. For a transition
from region I to region II to happen, $\dot{\tilde{K}}_{\varphi}>0$ when
$\delta\tilde{K}_{\varphi}=\frac{1}{2}\pi$, which by continuity extends to a
region around the transition hypersurface. Since $K_{\varphi}$ is a continuous
function of $\tilde{K}_{\varphi}$, it approaches the same value at the
transition hypersurface from both regions, given by $\delta
K_{\varphi}=1$. Applying (\ref{Can}), we see that the corresponding analog
solutions $K_{\varphi}$ behave like time reversed versions in a neighborhood
of the transition hypersurface: $\dot{K}_{\varphi}=\delta
\cos(\delta\tilde{K}_{\varphi}) \dot{\tilde{K}}_{\varphi}$ has opposite signs
on the two sides of the transition hypersurface because
$\cos(\delta\tilde{K}_{\varphi})$ has opposite signs in the two regions while
$\dot{\tilde{K}}_{\varphi}>0$ as we already saw.

For the same reason, $E^{\varphi}$ has opposite signs on the two sides and,
unlike $K_{\varphi}$, is not continuous because it goes through infinity if
$\tilde{E}^{\varphi}$ remains finite. (The classical equations of motion imply
that $K_{\varphi}$ is proportional to $\dot{E}^x$ rather than
$\dot{E}^{\varphi}$, such that it may remain regular while $E^{\varphi}$ grows
without bounds.) Therefore, the time derivative of the absolute value
$|E^{\varphi}|$, which is relevant for $q_{xx}$ in (\ref{qE}), has opposite
signs on the two sides: The second term in
\begin{equation}
 |E^{\varphi}|^{\bullet} = {\rm sgn}(E^{\varphi})\left(
 \frac{\dot{\tilde{E}}{}^{\varphi}}{\cos(\delta\tilde{K}_{\varphi})} + \delta
 \frac{\tilde{E}^{\varphi}}{\cos^2(\delta\tilde{K}_{\varphi})}
 \sin(\delta\tilde{K}_{\varphi}) \dot{\tilde{K}}_{\varphi}\right)\sim
 \delta \frac{{\rm sgn}(E^{\varphi})}{\cos^2(\delta\tilde{K}_{\varphi})}
 \tilde{E}^{\varphi} K_{\varphi} \dot{\tilde{K}}_{\varphi}
\end{equation}
is dominant near the hypersurface and has opposite signs on the two sides.
The geometry in region II can therefore be interpreted as a time-reversed
classical solution compared with the time direction in region I. (It is not
necessarily a time reversal of the same solution as in region I because
$E^{\varphi}$ is not continuous across the transition hypersurface.) 

Applying this result to all transitions, we see that a global solution of the
modified theory is a concatenation of infinitely many classical regions with
alternating orientations of time. In each region, the geometry is
indistinguishable from a classical solution. The only new physics therefore
resides in the time reversals, which make it possible that $K_{\varphi}$ can
remain bounded.

\subsection{Non-covariance}

In each local region, the geometry is covariant and slicing independent,
provided the changes of coordinates and slicings are sufficiently ``small''
such that they do not leave the range of $K_{\varphi}$ relevant for the
region. (We can apply slicing independence only in the classical analogs,
where the correct version (\ref{HH}) of hypersurface deformations holds.)
Globally, space-time in this model would be covariant only if the reversal
surfaces were covariantly defined, but this is not the case: They refer to
fixed values of $\delta K_{\varphi}=\pm 1$, and $K_{\varphi}$ is not a
space-time scalar. 

Choosing a different slicing in a classical analog in general shifts the
positions of time reversal surfaces. A complete solution for
$\tilde{K}_{\varphi}$ and $\tilde{E}^{\varphi}$ therefore violates slicing
independence, even after it has locally been mapped to a suitable effective
line element. For instance, in a vacuum solution there would be no time
reversals outside the horizon in a Schwarzschild slicing, but there are other
exterior slicings in which $\delta K_{\varphi}$ can be large and trigger time
reversal in the modified geometry. Even with minimal modifications introduced
by the model, general covariance is violated.

\section{Conclusions}

We have presented a detailed analysis of space-time structure in models
obtained by bijective or non-bijective canonical transformations of classical
gravity. Although the bijective case is completely equivalent to classical
gravity, a space-time analysis is non-trivial because the equivalence may be
hidden if complicated canonical transformations are applied. Our discussion
showed that basic fields of a modified theory, in general, cannot be
identified directly with metric components that play the same role as their
classical counterparts. 

While such a model would be considered trivial from the perspective of
modified gravity, it is nevertheless instructive because it highlights the
subtle nature of space-time structure in canonical theories. In particular,
the importance of identifying suitable metric components or effective line
elements constructed from the basic fields of a canonical modified theory
remains highly relevant if the theory is genuinely modified. The non-trivial
nature of such identifications has often been overlooked in models of loop
quantum gravity.

We applied our detailed construction of effective line elements that
consistently describe the space-time geometry of solutions in the modified
theory introduced in \cite{CovPol}. This model uses a non-bijective canonical
transformation and is therefore inequivalent to classical gravity.  However,
we have shown that the only new physical effect is the introduction of
time-reversal surfaces connecting classical space-time regions.  This
observation corrects the claim ``As [the canonical transformation] is
not-invertible in the whole of phase space it still allows to have the usual
novel phenomena that loop quantizations introduce in regions where one expects
general relativity not to be valid, like close to singularities.'' made in
\cite{CovPol}. Locally, general relativity is valid in all regions of the
modified theory, without any novel phenomena that have been claimed previously
in loop quantizations. Our constructions also show that effective geometries
described by the model depend only on the local maxima of the function
$K_{\varphi}(\tilde{K}_{\varphi})$. The specific sine function, usually
motivated by expressions of holonomies used in loop quantum gravity, does not
matter at all.

Even though the modifications are obtained by a canonical transformation of a
covariant theory, their global solutions violate covariance precisely at those
places where ``novel phenomena'' happen. This outcome heightens the covariance
crisis of loop quantum gravity: Even a minor modification of the classical
equations, inspired by loop quantum gravity but implemented by a canonical
transformation, is in conflict with the requirement of general covariance.

\section*{Acknowledgements}

The author thanks Rodolfo Gambini and Jorge Pullin for sharing a draft of
\cite{CovPol} and subsequent discussions. This work was supported in part by
NSF grant PHY-1912168.


\end{document}